\documentclass[10pt]{amsart}
\allowdisplaybreaks[4]
\usepackage{amssymb,color}
\usepackage{amsmath,amssymb,graphicx}
\usepackage{accents}
\usepackage{bbm}
\usepackage{lipsum}
\usepackage{array}
\usepackage{multirow}
\usepackage{tabularx,ragged2e,booktabs,caption}
\usepackage{here}
\usepackage[toc,page]{appendix}
\usepackage{adjustbox}
\usepackage[colorlinks=true, citecolor=blue, linkcolor=blue, urlcolor=blue ]{hyperref}

\usepackage{amsthm}
\usepackage{subcaption}
\usepackage[subnum]{cases}

\usepackage[numbers]{natbib}         

\usepackage{booktabs}
\usepackage{float}
\newcommand{\ra}[1]{\renewcommand{\arraystretch}{#1}}

\usepackage{amsthm}
\usepackage{subcaption}
\numberwithin{equation}{section}
\numberwithin{figure}{section}
\numberwithin{table}{section}
\allowdisplaybreaks[4]

\definecolor{c20}{rgb}{0.,0,0.}
\definecolor{c30}{rgb}{1,0.5,0.5}
\definecolor{c40}{rgb}{0,.7,0}
\definecolor{c50}{rgb}{1,0,1}

\def\TE#1{\textcolor{c30}{#1}}

\def\TE#1{#1}

\definecolor{c2}{rgb}{0,0.7,0.3}

\newtheorem{theo}{Theorem}[section]
\newtheorem{sat}[theo]{Proposition}
\newtheorem{de}[theo]{Definition}
\newtheorem{lem}[theo]{Lemma}
\newtheorem{exxa}[theo]{Example}

\newtheorem{korr}[theo]{Corollary}

\newtheorem{remarks}[theo]{Remarks}

\newcommand{\kb}[1]{\boldsymbol{#1}}
\newcommand{\vk}[1]{\kb{#1}}

\newcommand{\abs}[1]{\left\lvert #1 \right\rvert}

\def\E#1{\mathbb{E}\left \{#1 \right\}}

\newcommand{\R}{\mathbb{R}}

\newcommand{\BQN}{\begin{eqnarray}}
\newcommand{\EQN}{\end{eqnarray}}

\newcommand{\BQNY}{\begin{eqnarray*}}
\newcommand{\EQNY}{\end{eqnarray*}}

\newcommand{\BS}{\begin{sat}}
\newcommand{\ES}{\end{sat}}
\newcommand{\BT}{\begin{theo}}
\newcommand{\ET}{\end{theo}}
\newcommand{\BK}{\begin{korr}}
\newcommand{\EK}{\end{korr}}

\newcommand{\BD}{\begin{de}}
\newcommand{\ED}{\end{de}}

\newcommand{\BIT}{\begin{itemize}}
\newcommand{\EIT}{\end{itemize}}
\newcommand{\BDI}{\begin{description}}
\newcommand{\EDI}{\end{description}}

\newcommand{\BRM}{\begin{remarks}}
\newcommand{\ERM}{\end{remarks}}

\newcommand{\BTH}{\begin{theo}}
\newcommand{\ETH}{\end{theo}}
\newcommand{\BPR}{\begin{sat}}
\newcommand{\EPR}{\end{sat}}

\newcommand{\BEX}{\begin{exxa}}
\newcommand{\EEX}{\end{exxa}}

\newcommand{\BC}{\begin{cases}}
\newcommand{\EC}{\end{cases}}
\newcommand{\COM}[1]{}
\newcommand{\BL}{\begin{lem}}
\newcommand{\EL}{\end{lem}}

\def\bqn#1{ { \begin{eqnarray} #1 \end{eqnarray}}}

\begin{document}

\title[Price Optimisation for New Business]{Price Optimisation for New Business}

\author{Maissa Tamraz}
\address{Maissa Tamraz, Department of Actuarial Science, University of Lausanne, UNIL-Dorigny 1015 Lausanne, Switzerland}

\author{Yaming Yang}
\address{Yaming Yang, School of Finance, Nankai University, 300350, Tianjin, PR China,
and Department of Actuarial Science, University of Lausanne, UNIL-Dorigny 1015 Lausanne, Switzerland}

\bigskip

\date{\today}
 \maketitle

\bigskip
\begin{quote}
{\bf Abstract}:  This contribution is  concerned with price optimisation of the  new business for  a  non-life product. Due to high competition in the insurance market, non-life insurers are interested in increasing their conversion rates on new business based on some profit level. In this respect, we consider the competition in the market to model the probability of accepting an offer for a specific customer. We study two optimisation problems relevant for the insurer and present some algorithmic solutions for both continuous and discrete case. Finally, we provide some applications to a motor insurance dataset.
\end{quote}



\def\atR{$@ \mathcal{R} \ $}
\def\atRR{$@ \mathcal{R}$}

\section{Introduction}

Consider that in the insurance market $N$ customers are looking for an insurance coverage. There are $k+1$ different insurance companies that offer different premiums to each customer, say $j$th  customer
receives $k$ offers, i.e., $P_{ji}, i\le k$ is the premium offered by the $i$th insurance company.
Of course, we shall assume that all the offers are for the same coverage. Of interest here is the possibility of a premium optimisation approach for a given company operating in the market. We write, for notation simplicity, the premiums offered by that  company to the $j$th customer  as $P_j$ instead of say $P_{j1}.$
Let us consider a simple example. Suppose that $k=3$, so there are altogether four companies in the market.
The premiums offered by three of them are 500CHF, 520CHF, 522CHF, whereas the premium offered by the company in question that will perform an optimisation, say $l$, is $P_j=519$CHF. Assume that the total profit from the contract if the premium offered is 500CHF equals 40CHF. If instead of $P_j$ an optimal premium
$$P_j^*=P_j(1+\delta_j), \quad 1 \le j \le N$$ is offered, for instance  $\delta_j=-0.06$ then the contract is still profitable (with approximate profit of 16 CHF) and moreover,  by this offer the company is ranked first.
The chances for getting this customer are therefore high. Typically, insurance companies offer also premiums that are not profitable (those risks are cross-subsidised). Therefore lowering the premium is not always the right and optimal solution. So the decision related to which $\delta$ to choose for each contract strongly depends on
the strategy of the company.  For the customer $j$ let $I_j(P_j, \delta)$ denote the Bernoulli random variable which equals to 1, meaning that the customer accepts the contract for the premium $P_j(1+ \delta_j)$ with acceptance probability
$$ \pi_j(P_j,\delta_j) \in (0,1].$$
Each contract offered can be seen as an independent risk. Therefore, the total number of customers that join the company as new business is given by
$$\mathcal{N}(\vk{\delta})= \sum_{j=1}^N I_j(P_j, \delta_j).$$
Hereafter, in order to avoid trivialities  we shall assume that $\delta_j>-1$. \\
Consequently, the total premium volume of the new business $\mathcal{V}(\vk{\delta})$ (which is random) is given by
$$ \mathcal{V}(\vk{\delta})= \sum_{j=1}^{N}  I_j(P_j, \delta_j) P_j (1+ \delta_j).$$
Of interest for the insurance company is to maximize the expected premium volume, i.e., the objective function is
to maximize
\bqn{\E{\mathcal{V}(\vk{\delta})}= \sum_{j=1}^NP_j (1+ \delta_j) \pi(P_j, \delta_j)}
under some business constraints, for instance the expected number of new customers should not be below $a N$, i.e.,
\bqn{ \E{ \mathcal{N}(\vk{\delta})}= \sum_{j=1}^N \pi_j(P_j, \delta_j) \ge aN,
}
where $a\in (0,1)$ is a prespecified known constant. \COM{In the sequel, we present two optimisation problems of interest to the insurance company who will perform the optimisation. }\\
Price optimisation for new business has already been discussed in brief, see \cite{PONB} and for renewal business see \cite{hashorva2017some}
  for more details. However, in the literature, price optimisation is more focused on the regulations and ethical points of view, see \cite{PROA,PROB}.\\
This paper is structured as follows. In Section 2, we define the optimisation problems from the insurer's perspective. Section 3 is dedicated to the choice of the acquisition rate $\pi_j$ for each customer based on the competitors' price. Finally, Section 4 presents applications of the defined optimisation problems to a simulated data set.

\section{Optimisation Models} \label{section 3}
 Nowadays, insurers are interested in increasing their conversion rates on new business. This action leads as a result to an increase in the premium volume of the respective company.  Clearly, one simple solution is to lower the premiums of all customers looking to purchase an insurance coverage in the market on one hand and increase the premiums of the existing customers at renewal on the other hand. Eventhough this method might substantially increase the conversion rate of company $l$ and its expected premium volume, it does not represent the optimal solution as it does not differentiate between the different segments of customers in the market. Therefore, we use price optimisation  in order to avoid negative profit performance and adverse selection. Moreover, given the high competition in the market, insurance companies need to constantly monitor their position to maintain their reputation. In this respect, we shall consider the competitors' premiums in the optimisation setting. \\

In the sequel, we assume that we have k+1 insurance companies in the market, representing k competitors for company $l$ who will perform the optimisation.  Also, we assume that we have full information about the market, so the premiums of the $k$ competitors are known.
Hereafter, we shall  define two optimisation problems relevant for company $l$  assuming that N customers are looking to purchase an insurance coverage in the market.\\

\subsection{Maximise the expected premium volume}
Insurers are interested in maximising their premium volume as it is one of the main source of profit of an insurance company. However, they expect to have in their portfolio a certain number of new customers and this based on the strategy of the company. Thus, the optimisation problem can be formulated as such
\BQN\label{eq:obj1}
\begin{aligned}
&\underset{\vk{\delta}}{ \text{max }} \sum_{j=1}^{N} P_j(1+\delta_j) \pi_j(P_j,\delta_j),\\
&\text{subject to  } \frac{1}{N}\sum_{j=1}^{N} \pi_j(P_j,\delta_j) \leqslant \ell_1 , \\
&\text{and  } \frac{1}{N}\sum_{j=1}^{N} \pi_j(P_j,\delta_j) \geqslant \ell_2,\\
\end{aligned}
\EQN

where $\ell_1, \ell_2 < 1$  are 2 constants. For instance, $\ell_1$ and $\ell_2$ may denote the ratio of the expected number of customers to the total number of customers  of the cheapest and the most expensive company in the market respectively. However, in practice, the total numbers of customers in a certain insurance company is not known by other companies, hence $\ell_1$ and $\ell_2$ are set by the insurers based on their strategy. \\

\subsection{Maximise the expected number of new customers} The second objective function is concerned with the number of customers that the company is expected to get at the beginning of the period. Eventhough maximising the premium volume is important for insurers, they, nonetheless, are interested in acquiring a maximum number of new customers as they may profit the insurance company in the long run. Thus, one of the main goals of  insurers is to maximise the expected number of customers that may accept the offer. Hence, the optimisation problem can be formulated as follows 

\BQN\label{eq:obj2}
\begin{aligned}
&\underset{\vk{\delta}}{ \text{max }} \sum_{j=1}^{N}\pi_j(P_j,\delta_j),\\
&\text{subject to  } \sum_{j=1}^{N}P_j(1+\delta_j)  \pi_j(P_j,\delta_j) \leqslant C_1 , \\
&\text{and  } \sum_{j=1}^{N} P_j(1+\delta_j) \pi_j(P_j,\delta_j) \geqslant C_2,\\
\end{aligned}
\EQN

where $C_1$ and $C_2$ are two constants. Typically, in practice, the insurer would like to maintain the  reputation of the insurance company in terms of premium volume and thus stay relatively in the same position he was in the market before performing the optimisation. Therefore, $C_1$ and $C_2$ depend on the expected premium volume of the competitors. For instance, $C_1=\sum_{j=1}^{N} P_{n,j}\pi_{n,j}(P_{n,j},0)$ and $C_2=\sum_{j=1}^{N} P_{m,j}\pi_{m,j}(P_{m,j},0)$ where $m$ and $n$ denote the $m^{th}$ and $n^{th}$ competitors of company $l$ in the market with $C_1 > C_2$.

\COM{\textbf{Case 1:} We suppose that we have full information about the market. The premiums of the $k$ competitors are known. Thus, company $l$ will perform an optimisation based  on maximising the expected premium volume subject to some constraints on the number of new customers that we shall expect to have at the beginning of the period. In this setting, the objective function can be formulated as such:\\
\BQN\label{eq:optimprob}
\begin{aligned}
&\underset{\vk{\delta}}{ \text{max }} \sum_{j=1}^{N} P_j(1+\delta_j) \pi_j(P_j,\delta_j)\\
&\text{subject to  } \frac{1}{N}\sum_{j=1}^{N} \pi_j(P_j,\delta_j) \geqslant \ell \\
\end{aligned}
\EQN}
\BRM \hfill\\
\begin{enumerate}
\item In the optimisation setting, we assume that the competitors do not react to the premium change of company $l$ who performs the optimisation.
\item In the insurance sector, there are multiple competitors in the market. However, we assume that customers are looking for large, nationally known insurer compared to a less expensive local known insurer. In this respect, we consider that  10 insurance companies are competing in the market.
\item We assume that the change in premium $\vk{\delta}$ has an upper and lower bound and this based on the insurer strategy. For instance, the insurer doesn't want to be the cheapest in the market nor the most expensive one. This entails that, for $j \leq N$, $\delta_j \in (m_j,M_j)$ where $m_j, M_j \in (-1,1)$.\\
\end{enumerate}
\ERM

\section{Choices for  $\pi_j$}
The optimisation problems are quite similar to the ones defined in Hashorva et al. \cite{hashorva2017some} for the renewal business. However, the main difference lies in the choice of the probabilities $\pi_j$'s for customer $j$. Clearly, $\pi_j$'s are strongly dictated by how many companies are offering in the market, and how much is the premium difference. In the sequel, we discuss some possible tractable choices for $\pi_j$'s. \\

\subsection{ Step function for $\pi_j$} \label{a)} Suppose that there are $k$ other competitors in the market and their premiums for the $j$th customer are known. If the current rate of company $l$, who performs the optimisation, is below one of the competitors' rate, then an increase in the rate level might not lead to a decrease in policies. In this respect, we model the conversion rate for customer $j$ based on the competitors' premiums, more specifically with respect to the cheapest and  highest premiums observed in the market with respective probabilities  $c_1$  and $c_2$  where $c_1 > c_2$ and $c_1,c_2 \in (0,1)$ as follows
\BQN\label{eq:stepfunction}
\pi_j(P_j,\delta_j) = c_1+  (c_2-c_1)\frac{P_{j}(1+\delta_j)- \underset{1 \leqslant i \leqslant k}{ \text{min }}(P_{ji})}{\underset{1 \leqslant i \leqslant k}{ \text{max }}(P_{ji})-\underset{1 \leqslant i \leqslant k}{ \text{min }}(P_{ji})}, & \text{for } P_j(1+\delta_j) \in (A_j,B_j). \\\notag
\EQN
Clearly, $\pi_j$ is a piece-wise linear, decreasing step function where the jumps are dictated by the difference in premiums between two offers.
It should be noted that $A_j$ and $B_j$ are the jump points from one level to another. For simplicity, they are  defined as the arithmetic average between two premium offers.\\

 \COM{Also, for simplicity, we consider that the premiums are ordered. Thus, we denote by $P_1$ the cheapest premium offered to customer $j$, by $P_2$ the second cheapest up to $P_{k+1 }$ the most expensive premium.}

The corresponding shape of $\pi_j$  is realistic from a practical point of view as for some premium ranges the customers' behaviour is the same relative to the different offers.
The table below illustrates the latter. In this example, we consider four insurance companies and estimate the values of $\pi_j$ for some  premium ranges.

\begin{center}
\begin{table}[H]\centering
\ra{1}
\begin{tabular}{@{}ccccccccccccc@{}}\toprule
$P_{j}$ & 500 &515& 520 &522\\ \midrule
Range for $P_{j}$ & (495,507) &(507,517)&(517,521)&(521,525)\\ \midrule
$\pi_j$ & 0.75 &0.50&0.40 & 0.30\\
\bottomrule
\end{tabular}
\caption{Values of $\pi_j$ relative to customer $j$ based on the different premium ranges.}
\end{table}
\end{center}

By considering this expression for $\pi_j$,  the objective functions in (\ref{eq:obj1}) and (\ref{eq:obj2}) are non linear discontinuous functions. Several methods in the literature are discussed to solve non-linear optimisation problems, see \cite{boggs1995sequentialB,fletcher1963rapidly,frank1956algorithm}. However, these methods rely upon some assumptions on the objective function such as continuity, existence of derivatives, unimodality, etc. Therefore, in order to solve the optimisation problems at hand, we use the genetic algorithm method (abbreviated  GA) described in Appendix A. GA is a widely popular method when it comes to this type of objective function. It has been explored in many areas such as optimisation, operation, engineering, evolutionary biology, machine learning, etc., see \cite{mitchell1998introduction,reid1996genetic,shankar2010solving,garg2016hybrid} for more details. GA uses historical information to speculate on new search points with expected improved performance.\\

\textit{Discrete case for $\delta$.} Throughout the paper, we assume that $\delta$ is continuous and can take any values in the interval $(m,M)$ where $m, M \in (0,1)$. We shall investigate now the case where the change in premium $\delta$  can only take finite integer values from a discrete set for all customers. Let's consider the case where the competitors' premiums and company's $l$ premium do not change. As $\pi_j$ for $j \leq N$  depends  constantly on the competitors' offers for the coverage in question and the position of the latter  in the market, $\pi_j$ varies for each customer $j$  independently from $\delta_j$ and $P_j$. Thus, for illustration purposes,  we compute  the values of $\pi_j, j=1,2,3$ for three different customers  based on the different change in premium $\delta_i$ for $i=1,\ldots,9$.\\

\begin{center}
\begin{table}[H]\centering
\ra{1}
\begin{tabular}{@{}ccccccccccccc@{}}\toprule

$i $ &  1 & 2 & 3 & 4 & 5 & 6 & 7 & 8 & 9 \\ \midrule
$\delta_i$ & -20\% &  -15\% & -10\% & -5\% &  0\% &  5\% &  10\% &  15\% &  20\% \\\midrule
$\pi_1$ & 0.750 & 0.750 & 0.750  & 0.656 &0.536 &0.416 & 0.300 & 0.300 & 0.300 \\
$\pi_2$ & 0.750 & 0.750 & 0.723  & 0.613 &0.504 &0.394 & 0.300 & 0.300 & 0.300 \\
$\pi_3$ & 0.750 & 0.728 & 0.618  & 0.508 &0.398 &0.300 & 0.300 & 0.300 & 0.300 \\
\bottomrule
\end{tabular}
\caption{Values of $\pi$ for 3 different customers based on $\delta_i$ for $i=1,\ldots,9$.} \label{table:discreteset}
\end{table}
\end{center}

The discrete opimisation problem is solved using GA, see Appendix A for more description on the algorithm. \\

\subsection {Linear function for $\pi_j$.}\label{b)} The simplest choice for $\pi_j$  is by considering the continuous version of the step function defined above. Referring to \cite{hashorva2017some}, we assume that for each customer $j$, $$\pi_j(P_j,\delta_j)= \alpha_j +\beta_j \delta_j,$$ where $\alpha_j$ and $\beta_j$ are two constants to be estimated in applications. In this case, (\ref{eq:obj1}) is a quadratic optimisation problem subject to linear constraints, see \cite{markowitz1956optimization}.\\

\subsection{Logistic model for $\pi_j$.}\label{c)} The third choice is motivated by the logistic regression model where the logit function shall be used to model the conversion rate. The latter is popular in the literature for the modeling of probabilities such as the probability of renewal or lapses observed in an insurance portfolio as well as the probability of merger of non-life insurers, see \cite{ hashorva2017some, guillen2011logistic,meador1986probability}.  Its expression is given by
\BQN\label{eq:logit}
\pi_j(P_j,\delta_j)= \frac{1}{1+c_j^{-1}e^{-T_j \delta_j}},
\EQN
where $c_j$ and $T_j$ are two constants to be estimated in applications. $c_j$ includes the competition in the market and can be expressed in terms of $\pi_j$ before premium change as follows $$c_j=\frac{\pi_j(P_j,0)}{1-\pi_j(P_j,0)},$$ whereas $T_j < 0$ models the elasticity of customer $j$ relative to the change in premium $\delta_j$. The greater $|T_j|$ , the more elastic the customer $j$ is when purchasing an insurance policy. For this choice of $\pi_j$, the optimisation problems (\ref{eq:obj1}) and (\ref{eq:obj2}) are non-linear subject to non-linear constraints. We use the Sequential Quadratic Programming (SQP) method to solve this type of constrained optimisation problems. This method is very popular in the literature, see \cite{boggs1995sequentialB,nickel1984sequential}. It is an iterative method which solves a quadratic subproblem at each point iterate. The solution to the latter determines a step direction for the next point iterate. We refer to Appendix B for more details on the algorithm. \\

\section{Insurance Applications}
This section is dedicated to the application of price optimisation to insurance datasets. In this respect, we consider a simulated dataset describing the production of the motor line of business. The premiums are known and are assumed to be fair across the different segments of customers. Typically, in practice, auto-insurance markets are highly competitive. Insurers intensively compete on several factors such as price, quality of service, etc. Therefore, we consider that 10 leading insurers are competing in the market. Also, we assume that the premiums offered by the competitors are uniformly distributed around the company's $l$ premiums who is performing the optimisation. Based on some characteristics on the insured and the type of vehicle, an offer is made by the insurance company for the coverage in question. In the sequel, we shall  consider a heterogenous portfolio consisting of $n=1'000$ policyholders. Different premiums are offered to different segments of customers. Moreover, we note that for each customer the position of the competitors in the market change with respect to the premium charged and coverage. Therefore, we assume that for each offer the rank of the competitors and the company $l$ in question change. \\
The table below presents some statistics relative to the premiums offered by company $l$ and its competitors.

\begin{center}
\begin{table}[H]\centering
\ra{1.1}
\begin{tabular}{@{}ccccccccccccc@{}}\toprule
& \multicolumn{1}{c}{Initial Premium} & \phantom{abc}& \multicolumn{9}{c}{Competitors' Premiums} & \phantom{abc} \\
\cmidrule{2-2}\cmidrule{4-12}
&$P_0$  &&   $P_1$ & $P_2$ &$P_3$& $P_4$ & $P_5$ &$P_6$ &$P_7$&$P_8$&$P_9$\\ \midrule
 Mean  	 & 	1'204 	 && 	 1'163 	 & 	 1'166 	 & 	 1'167 	 & 	 1'164 	 & 	 1'172 	 & 	 1'168 	 & 	 1'164 	 & 	 1'163 	 & 	 1'162 	 \\ \midrule
 Min  	 & 	 400 	 && 	 371 	 & 	 374 	 & 	 381 	 & 	 369 	 & 	 367 	 & 	 373 	 & 	 376 	 & 	 368 	 & 	 368 	 \\
 Q1 	 & 	 804 	 && 	 792 	 & 	 789 	 & 	 797 	 & 	 795 	 & 	 800 	 & 	 795 	 & 	 801 	 & 	 792 	 & 	 799 	 \\
Q2 	 & 	 1'223 	 && 	 1'145 	 & 	 1'141 	 & 	 1'137 	 & 	 1'146 	 & 	 1'161 	 & 	 1'155 	 & 	 1'143 	 & 	 1'151 	 & 	 1'140 	 \\
 Q3 	 & 	 1'598 	 && 	 1'500 	 & 	 1'501 	 & 	 1'495 	 & 	 1'517 	 & 	 1'521 	 & 	 1'505 	 & 	 1'489 	 & 	 1'496 	 & 	 1'493 	 \\
 Max 	 & 	 1'999 	 && 	 2'248 	 & 	 2'190 	 & 	 2'233 	 & 	 2'252 	 & 	 2'233 	 & 	 2'235 	 & 	 2'176 	 & 	 2'217 	 & 	 2'217 	 \\
\bottomrule
\end{tabular}
\caption{Premium Statistics }\label{table:prem_stat}
\end{table}
\end{center}

The above statistics rely on a simulation procedure described under the following steps:
\begin{itemize}
\item \textbf{Step 1:} Generate n uniform random numbers between $400$ and $2'000$. These numbers account for the premiums offered by company $l$ for a given coverage, denoted by $P_0$. We assume that $75\%$ of the customers are charged a premium between $400$ and $1'600$ and the rest in $(1'600,2'000)$.  This assumption is accurate in practice especially for  \textbf{TPL covers and All Risks.}
\item \textbf{Step 2:} Simulate n uniform random numbers $u$  between $0.25$ and $0.75$. These numbers reflect the ranks of company $l$ relative to each offer. For instance, $0$ and $1$ are the ranks of the cheapest and the most expensive companies competing in the market respectively whereas say $0.5$ is the company ranked $5^{th}$ among the $10$ competitors.
\item \textbf{Step 3:} Based on \textbf{Step 2}, we compute the median premiums, denoted by $P_{m_j}$, for each offer $j$. We assume that the smallest and greatest premiums in the market, denoted by $P_{min_j}$ and $P_{max_j}$ are expressed with respect to $P_{m_j}$ as such
$$P_{min_j}= P_{m_j} (1+ lwb) \quad \text{and} \quad P_{max_j}=P_{m_j}(1+upb) \quad \text{with}\quad lwb=-10\% \text{ and } upb=15\%.$$
Hence, for each offer $j$, $P_{m_j}$ is computed as follows
\BQNY
\frac{P_{m_j}-P_{0j}}{0.5- u}=\frac{P_{max_j}-P_{min_j}}{1-0}  \implies P_{m_j}=\frac{P_{0j}}{1+(upb-lwb)(u-0.5)}.
\EQNY
\item \textbf{Step 4:} For $j \leq n$, we generate premiums between $(P_{min_j}, P_{max_j})$ with $P_{min_j}$ and $P_{max_j}$ as defined in \textbf{Step 3} for the remaining $7$ insurance companies denoted by $P_{ij}$ for $i=1,\ldots,7$.
\item \textbf{Step 5:} If $P_{ij}=P_{0j}$ for $i=1,\ldots,7$, go back to \textbf{Step 4.}\\
\end{itemize}

\subsection{$\pi_j$ as defined in \ref{a)}} We consider $\pi_j$ for customer $j$ to be a step function as defined in (\ref{eq:stepfunction}). Let $c_1=0.75$ and $c_2=0.3$ denote the conversion rates for the cheapest and the most expensive insurance companies offering in the market respectively.  These values are accurate from the perspective of the policyholder as the latter is not only interested in paying the lowest premium offered in the market but is also interested in the reputation of the company and the quality of service. For illustration purposes, we first assume that only  one customer is looking to purchase a motor insurance policy in the market. The competitors' offers are summarized below and are ranked in ascending order:
438, 457,  477,  492,  532,  596,  654, 675,  733.\\
Company $l$ who will perform the optimisation offers an initial premium of $P_0=  568 $ to the corresponding customer for the coverage.
The figure below shows the conversion rates based on the different premiums offered.\\
\begin{figure} [H]
	\centering
	\caption{ Values of $\pi_j$ based on the premium range for customer $j$. }
		\label{Figure: step_function}
	\end{figure}
For instance, if the insurer decides to increase the premium of the current policyholder from $568$ to say $680$, the probability of acquiring the new business decreases from $0.55$ to $0.4$. Whereas, a small increase in premium, say $580$,  will yield the same conversion rate.\\

In the sequel, we assume, for simplicity,  that the competitors do not react to increases/ decreases in premiums of company $l$. This assumption is accurate from a practical point of view as the reaction of the market to price changes is unlikely to be instantaneous due to several factors. One of the main factors is the time delay in settling claims. Indeed, the latter is important when  modeling  the financial state of a company. \\
Hereafter, we denote by $t_0$ the time before optimisation is performed and by $t_1$ the time after the optimisation. We consider the optimisation problems defined in Section \ref{section 3}.

\subsubsection{Maximise the expected premium volume at $t_1$} \hfill\\

\textit{i) Continuous case for $\delta$.} We consider the optimisation problem defined in (\ref{eq:obj1}). The objective of the insurer is to maximise the premium volume of the company under some constraints on the number of customers that he expects to get at the beginning of the insurance period.
In the sequel, we shall consider  a conversion rate between $45\%$ and $50\%$. The change in premium $\delta$ lies in $(-20\%, 20\%).$ The optimal results obtained when solving (\ref{eq:obj1}) using the function \textit{ga in Matlab} are presented hereafter.\\

Figure \ref{Figure: Premium_comp} below highlights the positions of company $l$ among its competitors  based on the premiums offered to $n=1'000$ customers looking to purchase an insurance coverage in the market at time $t_0$ and $t_1$.

\begin{figure} [H]
	\centering
	\caption{ Premiums offered by company $l$ compared to the competitors. }
		\label{Figure: Premium_comp}
	\end{figure}

Figure \ref{Figure: Premium_comp} shows that $4\%$ of the customers in the market are offered the highest premium by company $l$ at $t_0$ compared to $39\%$ after optimisation, at $t_1$. The percentage of customers that are offered the cheapest premium in the market at $t_1$ is of $29\%$ compared to none at $t_0$. This relatively high increase will generate new sales for company $l$. In this particular case, the premium charged is lower than the market average premium. In practice, this decrease in premium may mainly target young new drivers. Also, if, for instance,  the new customers' family, say parents or siblings, are already  insured  within the company, the decrease in premium is a way to enhance the loyalty of the individuals towards the company  and thus increasing their future lifetime within the company.  Finally, $32\%$ are offered a premium in between at $t_1$ compared to $96\%$ at $t_0$.  \\

In the sequel, we shall consider two scenarios with respect to different constraints and different bounds for $\delta$. \\
\textbf{Scenario 1:} The expected percentage of new customers  (abbreviated EPN)  shall be between $45\%$ and $50\%$.\\
\textbf{Scenario 2:} The  EPN shall be between $50\%$ and $55\%$.\\
The table below shows the optimal results at $t_1$ for the two scenarios. All optimal results are normalised with the results obtained from the assumption that the insurer will not change the premiums for next year.
\begin{center}
\begin{table}[H]\centering
\ra{1}
\begin{tabular}{@{}ccccccccccccc@{}}\toprule
& \multicolumn{2}{c}{Scenario 1} & \phantom{abc}& \multicolumn{2}{c}{Scenario 2} \\
\cmidrule{2-3} \cmidrule{5-6}
Bounds for $\delta$ &$(-10\%,15\%)$ & $(-20\%,20\%)$  && $ (-10\%,15\%)$ & $ (-20\%,20\%)$\\ \midrule
Aggregate expected future premium at t1 (\%)	&	107.51	&	106.41	&&	108.29	&	114.91	\\
Expected number of new policies at t1 (\%)	&	107.07	&	106.94	&&	109.22	&	117.74	\\
Average optimal $\delta$ (\%)	&	1.44	&	0.48	&&	0.81	&	-2.63	\\
Average optimal increase (\%)	&	13.80	&	13.78	&&	13.66	&	13.82	\\
Average optimal decrease (\%)	&	-9.34	&	-13.37	&&	-9.41	&	-13.14	\\
Number of increases	&	466	&	510	&&	443	&	390	\\
Number of decreases	&	534	&	490	&&	557	&	610	\\
\bottomrule
\end{tabular}
\caption{Optimal results for Scenario 1\&2  based on different bounds for $\delta$}\label{table:opt_results}
\end{table}
\end{center}

Table \ref{table:opt_results} shows that for both scenarios,  the average optimal $\delta$ decreases with the range of possible premium changes.  For instance, in Scenario 1, the average optimal $\delta$ decreases from $1.44\%$ for $\delta \in (-10\%,15\%)$ to $0.48\%$ for $\delta \in (-20\%,20\%).$
Also, the expected premium volume for Scenario 2 is greater than the one in Scenario 1 for both bounds. This is mainly due to the fact that the constraint on the expected number of new customers that will join company $l$ is greater in Scenario 2.  Typically, the probability that new customers will join the company is higher resulting in a positive effect on the expected premium volume. Also, the number of customers subject to an increase in premium is higher in Scenario 1 for both bounds.\\

\BRM
In this application, we considered a portfolio of $n=1'000$ customers  looking to purchase an insurance policy in the market. However, in practice,  n is larger. For n large, the running time may take hours (for instance, for $n=10'000$, the running time is about four hours) whereas for $n=1'000$ customers, using matlab, the running time is about $15$ min which is reasonable. Hence, it is less time consuming if the insurance company  split the offers into different categories and perform the optimisation for each of these different segments.
\ERM

\textit{ii) Discrete case for $\delta$.}
In the following, we consider that the change in premium $\delta$ takes its values from a discrete set as seen in Table \ref{table:discreteset}.  We look at the optimisation setting (\ref{eq:obj1}) and set the constraints on the expected conversion rate between $45\%$ and $50\%$.  To solve (\ref{eq:obj1}), the function \textit{ga} implemented in Matlab is used.  The optimal results are summarized below.\\

Figure \ref{Figure: Optimaldelta} highlights the distribution of the $1'000$ offers with respect to the set of discrete premium changes. For $93.5\%$ of the portfolio, $\delta$ is between $-15\%$ and $15\%$. Only a small proportion of customers are offered the lowest and highest change in premium, i.e. $-20\%$ and $20\%$.
\begin{figure} [H]
	\centering
	\caption{ Optimal change in premium for the whole portfolio . }
		\label{Figure: Optimaldelta}
	\end{figure}
\COM{\begin{figure} [H]
	\centering
	\includegraphics [scale=0.85]{Premium-compdisc.png}
	\caption{  Premiums offered by company $l$ compared to the competitors.. }
		\label{Figure: Comparison2}
	\end{figure}}

We consider two Scenarios based on the constraints of the optimisation problem.  \\
\textbf{Scenario 1:} The constraints on the EPN are: $\ell_1 =50\%$, $\ell_2=45\%$,\\
\textbf{Scenario 2:} The constraints on the EPN are: $\ell_1 =55\%$, $\ell_2=50\%$.\\

\begin{center}
\begin{table}[H]\centering
\ra{1}
\begin{tabular}{@{}ccccccccccccc@{}}\toprule
& \multicolumn{1}{c}{Scenario 1} & \phantom{abc}& \multicolumn{1}{c}{Scenario 2} \\
 \midrule
Aggregate expected future premium at t1 (\%)	&	106.80	&&	112.33	\\
Expected number of new policies at t1 (\%)	&	106.99	&&	117.65	\\
Average optimal $\delta$ (\%)	&	-0.31	&&	-2.85	\\
Average optimal increase (\%)	&	10.19	&&	16.16	\\
Average optimal decrease (\%)	&	-10.28	&&	-15.97	\\
Number of increases	&	 413 	&&	371	\\
Number of decreases	&	 587 	&&	629	\\

\bottomrule
\end{tabular}
\caption{Optimal results for Scenario 1\&2  based on different bounds for $\delta$}\label{table:optimalresults3}
\end{table}
\end{center}

\COM{\begin{center}
	\begin{tabular}{|c||c|c|}
\hline
Constraints on the conversion rate & \textbf{Scenario 1}  & \textbf{Scenario 2}\\
\hline
Growth in expected premium volume at t1& $14.24\%$ & $9.62\%$   \\
\hline
Growth in expected number of customers  at t1& $29.82\%$ & $17.00\%$ \\
\hline
Average optimal $\delta$ & $-2.43\%$ & $-0.28\%$\\
\hline
Number of increases& $478$ & $563$  \\
Average increase $\delta$ & $8.92\%$ &$6.81\%$ \\
\hline
Number of decreases & $522$ & $437$ \\
Average decrease $\delta$ & $-12.82\%$ & $-9.41\%$ \\
\hline
\end{tabular}
	\captionof{table}{Optimal results for Scenarios 1\&2. }
	\label{table:optimalresults3}
\end{center}}

Table \ref{table:optimalresults3} shows that for the same range of possible changes in premium $\delta$, an increase in the expected conversion rate leads to a higher expected premium volume and a lower average optimal $\delta$ as  seen in Scenario 1.\\

\subsubsection{Maximise the expected percentage of new business at $t_1$} \hfill \\
We shall now investigate the optimisation problem  (\ref{eq:obj2}) where the insurer maximises the expected number of new customers that will join company $l$. Clearly, in this case, a simple approach is to offer the lowest premium in the market to attract a maximum number of new customers. However, this is not beneficial to the insurer as he would like to maintain the reputation of the insurance company in the market. In this respect, let $C_1$ and $C_2$ be two constraints relative to the expected premium volume set by the insurer for the optimisation. We assume that $C_1$ and $C_2$ depend on the expected premium volume of the competitors.\\

In the sequel, we shall analyse two Scenarios.\\
\textbf{Scenario 1:} We assume that the growth in the expected premium volume is  between $8\%$ and $10\%$, i.e,  $C_1= 595,033$ and $C_2=583,130$.\\
\textbf{Scenario 2:} We assume that the growth in the expected premium volume is  between $10\%$ and $16\%$ , i.e, $C_1= 624,410$ and $C_2=595,033$ .\\

The figure below compares the number of customers observed in the different premium ranges along with the average optimal change $\delta$ in each range at time $t_0$ and $t_1$, i.e. before and after the optimisation  is performed, under both Scenarios.
\begin{figure} [H]
	\centering
	\caption{ Number of policyholders in each premium range at t0 and at t1 for both Scenarios. }
		\label{Figure: Comparison}
	\end{figure}
As seen in Figure \ref{Figure: Comparison},  the number of offers with a premium less than $600$ increases at $t_1$ whereas the number of offers with a premium above $1'800$ decreases under both Scenarios. This decrease in premium is explained by a negative average optimal change $\delta$  for almost all ranges of premiums. Also,  the curve of the average optimal $\delta$ for Scenario 1 is always above the one for Scenario 2, i.e, the average optimal $\delta$ increases at a faster pace in Scenario 1 compared to Scenario 2. This is mainly explained by the constraints on the expected premium volume which is higher in Scenario 2.

\begin{center}
\begin{table}[H]\centering
\ra{1}
\begin{tabular}{@{}ccccccccccccc@{}}\toprule
& \multicolumn{1}{c}{Scenario 1} & \phantom{abc}& \multicolumn{1}{c}{Scenario 2} \\
\cmidrule{2-2} \cmidrule{4-4}
Bounds for $\delta$ &$(-20\%,20\%)$ && $(-20\%,20\%)$\\ \midrule
Aggregate expected future premium at t1 (\%)	&	109.68	&&	115.10	\\
Expected number of new policies at t1 (\%)	&	118.00	&&	124.13	\\
Average optimal $\delta$ (\%)	&	-3.03	&&	-4.97	\\
Average optimal increase (\%)	&	16.30	&&	16.20	\\
Average optimal decrease (\%)	&	-16.46	&&	-16.02	\\
Number of increases	&	410	&&	343	\\
Number of decreases	&	590	&&	657	\\

\bottomrule
\end{tabular}
\caption{Optimal results for Scenario 1\&2  based on different bounds for $\delta$}\label{table:optimalresults2}
\end{table}
\end{center}

Table \ref{table:optimalresults2} shows that an increase in the expected premium volume leads to an increase in the expected number of customers and a decrease in the average optimal $\delta$. These results are accurate from the insurance company's perspective as the conversion rate increases when  $\delta$ decreases leading to a higher expected premium volume.\\

\subsection{$\pi_j$ defined as in Section  \ref{c)}}
In this section, we consider the logit function, as defined in (\ref{c)}), commonly used to model the elasticity of a customer due to price changes. In Hashorva et al.  \cite{hashorva2017some}, the latter was used to model the probability of renewal after premium change and in \cite{guillen2011logistic}, the lapses observed in the insurance industry. We assume that the constraint on the expected percentage of customers to accept the offer of company $l$ is of $45\%$ and the change in premium $\delta$ lies in the interval $(-20\%,20\%)$. Solving the optimisation problem defined in (\ref{eq:obj1}) yields the following optimal results. First, the figure below shows the ranks of company $l$ among the competitors based on the premium offers to the n customers before and after performing the optimisation.

\begin{figure}[!htb]
   \begin{minipage}{0.50\textwidth}
     \centering
     \caption{Acquisition rate: 45\%}\label{Figure: logit1}
   \end{minipage}\hfill
   \begin {minipage}{0.50\textwidth}
     \centering
     \caption{Acquisition rate: 50\%}\label{Figure: logit2}
   \end{minipage}
\end{figure}

\COM{\begin{figure} [H]
	\centering
	\includegraphics [scale=0.80]{newoptlogit.png}
	\caption{ Number of policyholders in each premium range at t0 and at t1 for both Scenarios. }
		\label{Figure: logit}
	\end{figure}}

Figure \ref{Figure: logit1} shows that a large proportion of customers are offered the highest premium in the market. This is mainly due to a relatively low acquisition rate of $45\%$.  Clearly, this is not the case when we increase the conversion rate. As seen in  Figure \ref{Figure: logit2}, a smaller proportion of customers are offered the highest premium in the market and a larger proportion are offered the smallest premium. This is accurate from the insurance company's perspective as the latter is more interested in acquiring new business than maximising its premium volume. \\

The table below summarizes the optimal results obtained at $t_1$ for the following two Scenarios based on different bounds for $\delta$.\\
\textbf{Scenario 1:} The constraint on the EPN is of $45\%$.\\
\textbf{Scenario 2:} The constraint on the EPN is of $50\%$.

\begin{center}
\begin{table}[H]\centering
\ra{1}
\begin{tabular}{@{}ccccccccccccc@{}}\toprule
& \multicolumn{2}{c}{Scenario 1} & \phantom{abc}& \multicolumn{2}{c}{Scenario 2} \\
\cmidrule{2-3} \cmidrule{5-6}
Bounds for $\delta$ &$(-15\%,15\%)$ & $(-20\%,20\%)$  && $ (-15\%,15\%)$ & $ (-20\%,20\%)$\\ \midrule
Aggregate expected future premium at t1 (\%)	&	106.36	&	108.03	&&	97.58	&	101.40	\\
Expected number of new policies at t1 (\%)	&	96.19	&	96.18	&&	106.86	&	106.86	\\
Average optimal $\delta$ (\%)	&	10.67	&	12.65	&&	-8.90	&	-4.82	\\
Average optimal increase (\%)	&	13.31	&	17.27	&&	14.31	&	18.18	\\
Average optimal decrease (\%)	&	-6.15	&	-8.83	&&	-14.81	&	-18.62	\\
Number of increases	&	864	&	823	&&	203	&	375	\\
Number of decreases	&	136	&	177	&&	797	&	625	\\

\bottomrule
\end{tabular}
\caption{Optimal results for Scenario 1\&2  based on different bounds for $\delta$}\label{table:opt_results2}
\end{table}
\end{center}

As seen in Table \ref{table:opt_results2}, Scenario 1 yields a higher expected premium volume for both bounds compared to Scenario 2 but results in a lower acquisition rate. The average optimal change in premium is positive in Scenario 1 whereas in Scenario 2 it is negative for both bounds. This is mainly explained by the acquisition rate set in both Scenarios. Finally, the larger the range of premium change, the higher the expected premium volume and the higher the average optimal $\delta$ for both Scenarios.

\section{Appendix A: Solution of (\ref{eq:obj1}) and (\ref{eq:obj2}) using (\ref{eq:stepfunction}).}

In this section, we illustrate the GA method used to solve the optimisation problem (\ref{eq:obj1}) with $\pi_j$ as defined in (\ref{eq:stepfunction}).
The constrained optimisation (\ref{eq:obj1}) can be reformulated as an unconstrained one by the penalty method as follows, see \cite{mitchell1998introduction}
\BQN\label{eq:unconstopt}
\begin{aligned}
&\underset{\vk{\delta}}{ \text{min }} f(\vk{\delta}),  \text { where }  f(\vk{\delta})= g(\vk{\delta})+ r \Bigl(\Phi[h_1(\vk{\delta})]+ \Phi[h_2(\vk{\delta}]\Bigr),
\end{aligned}
\EQN

where
$$
\begin{cases}
g(\vk{\delta})=\sum_{j=1}^{N}P_j(1+\delta_j)  \pi_j(P_j,\delta_j) ,\\
h_1(\vk{\delta})= -\sum_{j=1}^{N}\pi_j(P_j,\delta_j)+\ell_1,\\
h_2(\vk{\delta})= \sum_{j=1}^{N}\pi_j(P_j,\delta_j)-\ell_2 ,\\
\end{cases}
$$
and $\Phi$ and $r$ are the penalty function and the penalty coefficient respectively.\\
The penalty function $\Phi$ penalize infeasible solutions (solutions that will not satisfy the constraints of the optimisation problem) by reducing their values in the objective function, thus favoring feasible solutions in the selection process, see \cite{yeniay2005penalty}. In most cases, $ \Phi[g(x)]=g(x)^2$ and the solution (\ref{eq:unconstopt}) tends to a feasible solution of  (\ref{eq:obj1}) when r is large.\\

The algorithm is described under the following steps.\\
\textbf{Step 1:}  Set a maximum number of generation $n_{max}$.\\

\textbf{Step 2:} Select an initial set of solution estimates chosen randomly (in contrast with the SQP method described below where one estimation point for $\vk{\delta}$ is needed). In the sequel, the set and the solution estimates are referred to as initial population and members respectively.\\

\textbf{Step 3:}  The GA method relies on three operators in the following order:
\begin{enumerate}
\item Reproduction: This function consists in reproducing copies of the members of the initial population according to the value of their objective function. Typically, the members with the highest value of the objective function, i.e. $|g(\vk{\delta})|$, have a higher probability in contributing to the next generation. This operator can be seen as a biased roulette wheel where each member has a roulette wheel slot proportional to the value of its objective function. Therefore, to reproduce, we spin the roulette $p_s$ times, where $p_s$ is the size of the initial population, in order to get the new members of the next generation.
\item Crossover: It follows the reproduction one. This operator mainly produces new members of the new generation by mating and swapping each pair of strings of two members from the new generation at random with probability $p_c$.
\item Mutation: This operator alterates randomly the position of two or more  values  for the same member with probability $p_m$ (very small in general).\\
\end{enumerate}

\textbf{Step 4:} The algorithm stops when the maximum number of generation is attained. The optimal $\vk{\delta}$ is chosen based on the highest value of its objective function $|g(\vk{\delta})|$.
\BRM \TE{i) It should be noted that GA operates on a coding of the solution estimates in the form of strings often chosen to be a concatenation of binary representation.\\
ii) To guarantee the success of the method, the crossover and mutations operators play an important role in finding the optimal solution as they generate new solution estimates and remove the less desired ones.\\
iii) The linear constraints and the upper and lower bounds  are satisfied throughout the optimisation.\\
iv) The optimisation problem (\ref{eq:obj2}) is sovled in a similar way.
}
\ERM

\section{Appendix B: Solution of (\ref{eq:obj1}) and (\ref{eq:obj2}) using (\ref{eq:logit})}

(\ref{eq:obj1}) can be reformulated as follows:

\BQN\label{eq:sqpobj}
\begin{aligned}
&\underset{\vk{\delta}}{ \text{min }} g(\vk{\delta}),\\
&\text{subject to  }h_1(\vk{\delta}) \leqslant 0, \quad h_2 (\vk{\delta}) \leqslant 0\\
&\text{and  } f_1(\delta_j) \leqslant 0, \quad f_2(\delta_j) \leqslant 0 \quad \text{ for } j\leq N.\\
\end{aligned}
\EQN
where $g$, $h_1$ and $h_2$ are defined in (\ref{eq:unconstopt}) and $f_1(\delta_j)= \delta_j-M_j$, $f_2(\delta_j)=-\delta_j+m_j$ for $j \leq N$.\\

\textbf{Step 1:} The Langrangian function relative to (\ref{eq:sqpobj}) is given by
$$\mathcal{L}:=\mathcal{L}(\vk{\delta},\lambda,\beta,\vk{\mu},\vk{\gamma})=g(\vk{\delta})+\lambda h_1(\vk{\delta})+ \beta h_2(\vk{\delta})+\sum_{j=1}^N \mu_j f_1(\delta_j)+ \sum_{j=1}^N \gamma_j f_2(\delta_j),$$
where $\lambda, \alpha \in \R$ and $\vk{\mu}, \vk{\gamma} \in \R^N$ are the Lagrangian multipliers.\\
Let ($\vk{\delta_0}$, $\lambda_0$, $\beta_0$, $\vk{\mu_0}$, $\vk{\gamma_0}$) be an estimate of the solution at $t=0$.\\

\textbf{Step 2:}  The SQP is an iterative process. Therefore, we define the next point iterate at  t+1 as follows
 $$\vk{\delta_{t+1}} =\vk{\delta_{t}}+\alpha \vk{s_{t}} \quad \text{ for } t \geq 0,$$
 where $\alpha \in (0,1)$ is the step length  and $\vk{s_t}$ is a step vector. \\
$\vk{s_t}:=(\vk{s_{t}^{\delta}}, s_{t}^{\lambda}, s_{t}^{\beta}, \vk{s_{t}^{\mu}},\vk{s_{t}^{\gamma}})$  shall solve the following quadratic sub-problem evaluated at
$(\vk{\delta_{t}}, \lambda_{t}, \beta_{t}, \vk{\mu_{t}}, \vk{\gamma_{t}})$ and defined as follows
\begin{equation}\label{eq:subQP}
\begin{aligned}
&\underset{\vk{s_t}}{\text{min}}\frac{1}{2}\vk{s_t}^\top  Q \vk{s_t} +\nabla g(\vk{\delta_{t}})^\top \mathbf{\vk{s_t}},\\
&\text{subject to }\\
& \nabla h_1(\vk{\delta_{t}})^\top \vk{s_t}+ h_1(\vk{\delta_{t}})\leq 0,\\
& \nabla h_2(\vk{\delta_{t}})^\top \vk{s_t}+ h_2(\vk{\delta_{t}})\leq 0,\\
& \nabla f_1(\delta_{{t},i})^\top \vk{s_t}+ f_1(\delta_{{t},i})\leq 0 \text{ for } i \le N,\\
&\nabla f_2(\delta_{{t},i})^\top \vk{s_t}+ f_2(\delta_{{t},i})\leq 0 \text{ for } i \le N.
\end{aligned}
\end{equation}
$Q$ is an approximation of the Hessian matrix of $\mathcal{L}$ updated at each iteration by the BFGS quazi Newton formula, $\nabla g$ the gradient of the objective function and $\nabla h_1$, $\nabla h_2$ , $\nabla f_1 $ and $\nabla f_2 $ the gradient of the constraint functions.\\
To guarantee the existence and uniqueness of a solution at each iteration, $Q$ maintains its sparsity and positive definetness properties.\\

Whereas, $\alpha$ is chosen as such
$$\phi (\vk{\delta_t}+\alpha \vk{s_t}) \leq \phi(\vk {\delta_t}),$$
with $\phi$ a merit function whose role is to ensure  convergence of the SQP method to a global solution after each iteration. $\phi$ is given by
 $$ \phi (x)= g(x)+r \Bigl( h_1(x) + h_2(x) + \sum_{j=1}^N f_1(x_j)+\sum_{j=1}^N f_2(x_j) \Bigr) \text{  and  }   r>\underset{1 \le i\le N}{\text{max}}(|\lambda|,|\beta|,|\mu_i|,|\gamma_i|).$$\\

\textbf{Step 3:} If the new point iterate satisfies the KKT conditions defined in Remark \ref{KKT}, then it is a local minimum and the SQP converges to that point.
If not, set $t=t+1$ and go back to \textbf {Step 2}.

\BRM\label{KKT}
 The success of the SQP algorithm is governed by the KKT conditions. Typically, if the point iterate doesn't satisfy these conditions, the SQP algorithm do not converge to the local/global optima. Thus, we shall define the KKT conditions as follows
\BQN \label{Pr2} \left\{
			 	\begin{array}{lcl}
\nabla \mathcal{L}  =  \vk{0}, \\
\lambda^* h_1(\vk{\delta^*})=  0,\\
\beta^* h_2(\vk{\delta^*}) = 0,\\
\mu_{i} f_1(\delta_j^*) =0  \text{ for  } j \le N, \\
\gamma_j f_2(\delta_j^*) =0  \text{ for  } j \le N, \\
h_1(\vk{\delta^*}) \le   0, \text{ } h_2(\vk{\delta^*}) \le   0, \\
f_1(\delta_j^*) \le  0,\text{ } f_2(\delta_j^*)  \le  0  \text{ for  } j\le N  \\
\lambda \ge  0, \quad \beta \ge  0, \quad  \vk{\mu} \ge \vk{0}, \quad \vk{\gamma} \ge \vk{0}.
\end{array}
      \right.
\EQN
\ERM

\bibliographystyle{ieeetr}

\bibliography{PO}

\end{document}